\documentclass{WileyMSP-template}
\usepackage{amsmath}
\usepackage{multirow}
\usepackage{todonotes}
\usepackage{hyperref} 
\hypersetup{
	hidelinks,
	colorlinks,
	breaklinks=true,
	urlcolor=blue,
	citecolor=blue,
	linkcolor=blue,
	bookmarksopen=false
}

\begin{document}

\pagestyle{fancy}
\rhead{\newline Article }

\title{Side-Chain Engineering Controls Molecular Packing and Vertical Electron Transport in Discrete n-Type NDI–T2 Oligomers}

\maketitle


\author{Zihao Liu},
\author{Hajar Komekoz},
\author{Nisarg Trivedi},
\author{Max Dorfner},
\author{Lokesh Rasabathina},
\author{Rico Ehrler},
\author{Dr. Mike Hambsch},
\author{Dr. Chen Wang},
\author{Prof. Dietrich R. T. Zahn},
\author{Prof. Jörg Matysik},
\author{Stefan C. B. Mannsfeld},
\author{Carsten Deibel},
\author{Olav Hellwig},
\author{Frank Ortmann},
\author{Michael Sommer},
\author{Georgeta Salvan}*



\begin{affiliations}

Zihao Liu, Prof. Georgeta Salvan\\
Magnetooptics, Institute of Physics, Chemnitz University of Technology, Chemnitz, Germany
Research Center for Materials, Architectures and Integration of Nanomembranes, Chemnitz University of Technology (MAIN), Chemnitz, Germany

Prof. Dietrich R. T. Zahn\\
Semiconductor Physics, Institute of Physics, Chemnitz University of Technology, Chemnitz, Germany
Research Center for Materials, Architectures and Integration of Nanomembranes, Chemnitz University of Technology (MAIN), Chemnitz, Germany

Hajar Komekoz, Prof. Michael Sommer\\
Polymer Chemistry, Institute of Chemistry, Chemnitz University of Technology, Chemnitz, Germany
Research Center for Materials, Architectures and Integration of Nanomembranes, Chemnitz University of Technology (MAIN), Chemnitz, Germany

Nisarg Trivedi, Max Dorfner\\
TUM School of Natural Sciences, Technical University of Munich, Garching b. München, Germany

Prof. Frank Ortmann\\
TUM School of Natural Sciences, Technical University of Munich, Garching b. München, Germany
Atomistic Modelling Center, Munich Data Science Institute, Technical University of Munich, Garching b. München, Germany

Lokesh Rasabathina, Rico Ehrler, Prof. Olav Hellwig\\
Magnetic Functional Materials, Institute of Physics,  Chemnitz University of Technology, Chemnitz, Germany
Research Center for Materials, Architectures and Integration of Nanomembranes, Chemnitz University of Technology (MAIN), Chemnitz, Germany

Mike Hambsch, Prof. Stefan C. B. Mannsfeld\\
Faculty of Electrical and Computer Engineering and Center for Advancing Electronics Dresden (cfaed), TUD Dresden University of Technology, 01062 Dresden, Germany

Prof. Jörg Matysik\\
Institute of Analytical Chemistry, Leipzig University, 04103 Leipzig, Germany

Chen Wang, Prof. Carsten Deibel\\
Optics and Photonics of Condensed Matter, Institute of Physics, Chemnitz University of Technology, Chemnitz, Germany

Email Address:

Prof. Georgeta Salvan\\
salvan@physik.tu-chemnitz.de

\end{affiliations}



\begin{abstract}

High-efficiency vertical charge transport in multilayer organic optoelectronic devices highly depends on molecular structure and arrangement.Using two T2-(NDI-T2)$_2$ conjugated oligomers with ethylhexyl (EH) and octyldodecyl (OD) side chains as a model system, their effects on electron transport are investigated using, among other techniques, atomic force microscopy (AFM), conductive atomic force microscopy (C-AFM), grazing-incidence wide-angle X-ray scattering (GIWAXS), steady-state optical spectroscopy, and time-dependent density functional theory (TDDFT) calculations. The oligomers of both materials exhibit relatively disordered structures in the as-cast state, whereas annealing enhances structural order and leads to side-chain-dependent changes in molecular stacking. EH-based films show a smaller annealing-induced photoluminescence shift than OD-based films, suggesting less pronounced stacking and conformational rearrangements, consistent with TDDFT-supported exciton analysis. The current-voltage characteristics are analyzed using the space-charge-limited current (SCLC) model. After annealing, both materials exhibit improved vertical electron mobility. The EH-based films reach a maximum mobility of $1.2 \times 10^{-4}\ \mathrm{cm^2,V^{-1},s^{-1}}$.The combined results suggest that the shorter EH side chains modify the out-of-plane packing environment and facilitate vertical electron transport. These results establish a structure–property relationship between side-chain structure, molecular order, exciton coupling and vertical mobility, providing a molecular design basis for optimising organic semiconductors through side-chain engineering.

\end{abstract}


\section*{Introduction}
Organic semiconductors have attracted significant attention as active materials in flexible electronic devices because their optoelectronic properties can be tuned through molecular design and they are suitable for solution processing~\cite{flex00,flex01,flex02,flex03,flex04,flex05,flex06}. In multilayer devices such as organic solar cells, organic light-emitting diodes, and organic spin valves, charge carriers must cross the electrode/organic layer interface to be injected and travel through the organic layer in a direction perpendicular to the substrate~\cite{flex06,osc01,photodetector01,LED01,xiong2004}. Therefore, device performance depends not only on interfacial energy level matching but also on vertical charge transport within the organic layers, the latter of which is closely related to molecular packing and structural order~\cite{Elvaligment01,Elvaligment02,wellordermolecular,keum2018}. However, existing research has primarily focused on in-plane charge transport in organic field-effect transistors, and there remains a lack of systematic understanding of how molecular structure and processing conditions jointly regulate out-of-plane transport~\cite{hidayat2021,gicevivcius2025}. This issue is particularly important for n-type organic semiconductors, as molecular structure and processing conditions jointly influence film formation and molecular packing, thereby modulating intermolecular electron coupling and vertical electron transport pathways~\cite{hidayat2021,yu2021}. Therefore, investigating the relationship among molecular structure, film packing, and vertical electron transport in n-type organic semiconductors is critical for optimizing the performance of vertical organic electronic devices.

Among n-type organic semiconductors, donor–acceptor conjugated systems based on naphthalene diimide (NDI) and bithiophene (T2) exhibit great potential because the electron-deficient NDI unit provides high electron affinity, while the NDI–T2 backbone enables efficient electron transport and good thermal stability~\cite{guo2014, sommer2014}. Charge transport in organic semiconductors is governed by the interplay between electronic coupling, molecular packing, energetic disorder, and molecular vibrations and may range from localized hopping or polaronic motion to transiently delocalized or more band-like regimes~\cite{X0,X1}. In conjugated materials, the backbone provides efficient intramolecular transport pathways, whereas long-range transport additionally requires intermolecular electronic coupling. Molecular packing therefore determines both the magnitude and directionality of charge transport~\cite{X2,vu2023}. Therefore, the microcrystalline molecular packing in the film strongly influences in-plane/out-of-plane charge mobilities and the anisotropic transport behavior. Although the benchmark polymer P(NDI2OD-T2) has been widely studied~\cite{N2200ElectronMobility, PDNIOD02,schuettfort2013,mohapatra2025,bucella2015,T2struc}, the batch-to-batch and molecular weight variation make studying the intrinsic structure-property relationship rather challenging~\cite{nahid2017unconventional,matsidik2016effects}.
In contrast, discrete conjugated oligomers offer an intermediate molecular scale between monomers and polymers. Compared with monomers, they provide a longer and well-defined conjugated backbone over which charge transport can be examined across several covalently connected units, while still retaining a precisely defined molecular structure. At the same time, unlike conventional polymers, their fixed chain length avoids complications arising from molecular-weight distribution and dispersity. These characteristics make conjugated oligomers particularly suitable for establishing clear correlations between molecular structure, packing, orientation, and charge transport~\cite{X3,X5}. Previously, we demonstrated synthetic access to discrete oligomers of the type T2-(NDI-T2) with chain lengths up to the heptamer~\cite{T2molecular}, which are ideal for studying transport. Most existing research on side-chain engineering has focused on either in-plane carrier transport in field-effect transistors or active-layer morphology and charge generation in bulk-heterojunction solar cells~\cite{sidechainPei,sidechain03,sidechain04, sidechain05}. However, there remains a lack of quantitative studies on out-of-plane carrier transport in n-type oligomers that integrate vertical electrical measurements, fine-structure characterization, and modeling of transport mechanisms.

In this work, two T2--(NDI--T2)$_2$ oligomers containing the same conjugated backbone but different branched side chains, octyldodecyl (OD) and 2-ethylhexyl (EH), were investigated. Thermal annealing near $T_{\mathrm{c}}$ was used to tune molecular packing and test its relation to the optical response and vertical charge transport. The resulting structural changes were examined using grazing-incidence wide-angle X-ray scattering (GIWAXS), X-ray reflectivity (XRR), and in-plane X-ray diffraction (IP-XRD), while film thickness and surface morphology were characterized by spectroscopic ellipsometry and atomic force microscopy (AFM). Changes in the optical response were evaluated using ultraviolet--visible (UV-vis) absorption and photoluminescence (PL) spectroscopy, supported by time-dependent density functional (TDDFT) theory calculations. Vertical electron transport was investigated at both the nanoscale and device level using conductive AFM and vertical electron-only diodes. By combining these measurements, correlations among side-chain structure, annealing-induced molecular packing, optical response, and out-of-plane electron transport were established, providing insights into the optimization of n-type organic semiconductor films for vertical electronic devices.

\section{Results and discussion} 

\subsection{Side-chain and annealing-driven differences in film Morphology}

\begin{figure}[!h]
  \includegraphics[width=\linewidth]{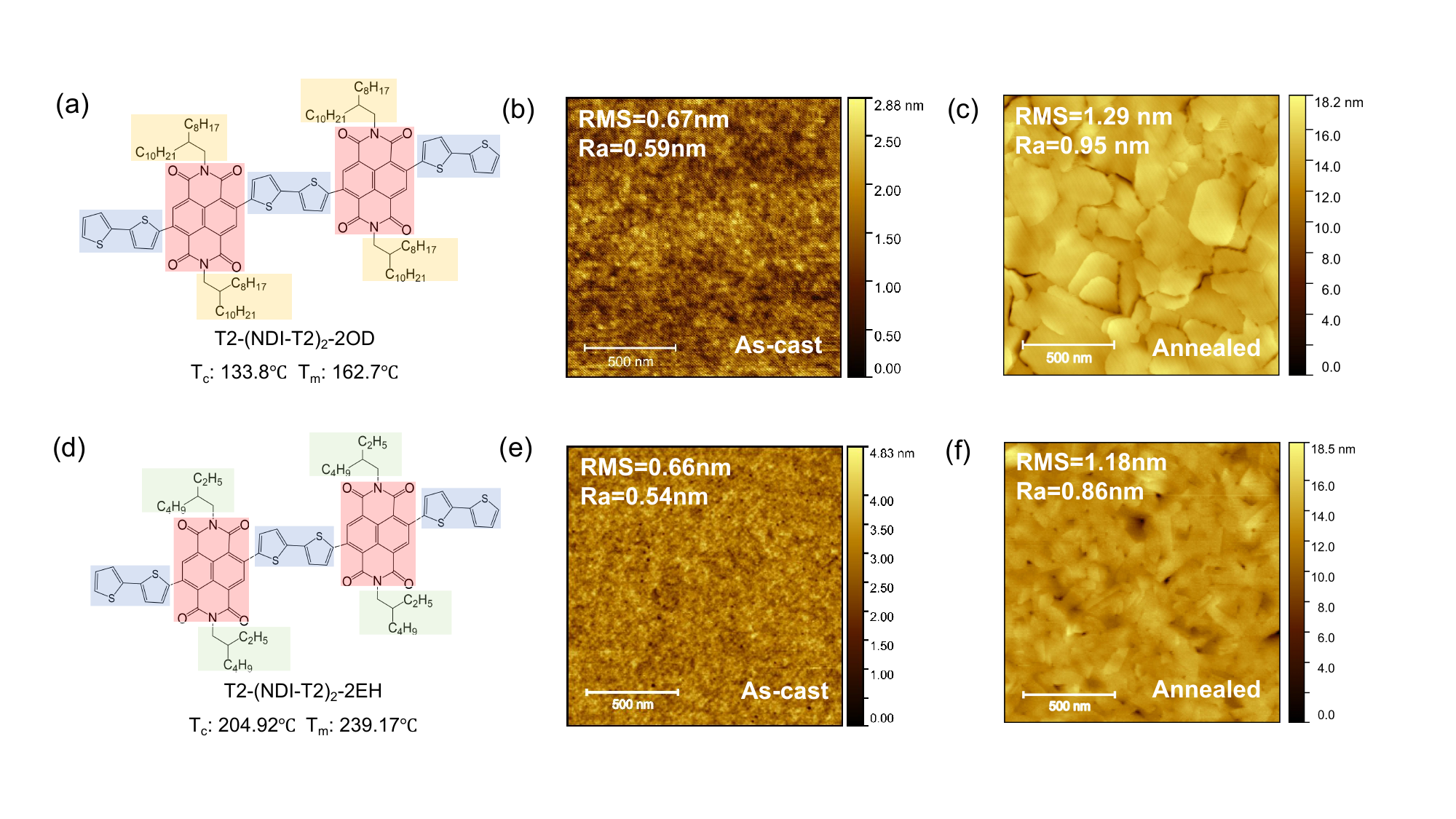}
 \caption{
(a,d) Chemical structures of the molecules investigated in this study. The blue, red, and yellow/green regions represent the donor, acceptor, and side-chain units, respectively. (b,c,e,f) AFM topography images of films prepared at 3000~rpm: (b,e) as-cast films and (c,f) films annealed for 30~min at 145~$^\circ$C for the OD-based material and 208~$^\circ$C for the EH-based material. Panels (a--c) correspond to T$_2$-(NDI-T$_2$)$_2$-2OD, while panels (d--f) correspond to T$_2$-(NDI-T$_2$)$_2$-2EH.
}
  \label{AFM}
\end{figure}

 T2-(NDI-T2)$_2$-2OD was synthesized according to the previously reported procedure~\cite{T2molecular}. The corresponding 2-ethylhexyl-substituted oligomer, T2-(NDI-T2)$_2$-2EH, was prepared by adapting the same direct-arylation strategy. Their molecular structures are shown in Figure~\ref{AFM}(a) and (d). For the sake of simplicity, the two oligomers will be labeled as OD and EH, respectively. To investigate the influence of side-chain structure on film properties, OD and EH films were prepared over a spin-coating speed range of 1,000 to 5,000 rpm. For the C-AFM sample series, the films were deposited onto Au (50\,nm)/Ta (4\,nm)/SiO$_2$ (100\,nm)/Si substrates, with OD film thicknesses ranging from approximately 8 to 18\,nm and EH film thicknesses ranging from approximately 7 to 14\,nm. The effect of spin-coating speed on film thickness and the corresponding calibration procedure are detailed in Section~S3 of the Supplementary Information (SI). By keeping the organic layer thicknesses comparable, the observed differences in final device performance can be directly attributed to changes in optical, morphological, and optoelectronic properties caused by molecular structure. For single-carrier devices, separate organic layers were fabricated on ZnO/ITO/glass substrates; their thickness data are provided in Section~S6 of the Supplementary Information (SI). This complementary set of samples enables the study of side-chain-dependent morphology and vertical transport characteristics under corresponding measurement conditions.

Differential scanning calorimetry (DSC)  measurements (shown in Figure S1 of the supporting information) revealed that the EH derivative exhibits substantially higher transition temperatures for both melting and crystallization processes ($T_\mathrm{c} = 204.9~^\circ$C, $T_\mathrm{m} = 239.2~^\circ$C) than the OD derivative ($T_\mathrm{c} = 133.8~^\circ$C, $T_\mathrm{m} = 162.7~^\circ$C). This phenomenon can be attributed to the distinct molecular structures of the side chains. Compared with the longer and more flexible OD side chain, the shorter EH side chain facilitates more compact molecular packing and stronger intermolecular interactions. For the annealing experiments throughout this work, the crystallization temperatures obtained from DSC were used as reference points. The annealing temperatures were chosen slightly above the respective $T_\mathrm{c}$ values to promote crystallization, resulting in 145~$^\circ$C for the OD-based and 208~$^\circ$C for the EH-based thin films and devices. The selected temperatures were not defined by a fixed offset from $T_\mathrm{c}$, but rather as practical annealing conditions above the crystallization transition for each material. We found that an annealing time of 30 min provides the best compromise between the crystallization of the film and a low number of pinholes between the crystalline domains, which is essential for avoiding short circuits in the vertical devices. This is why all annealing experiments presented in this study were performed for 30 min. 

To investigate the effects of side-chain structure and thermal annealing on the film morphology, atomic force microscopy (AFM) measurements were performed. Before annealing, both OD- and EH-based films exhibited relatively uniform surface morphologies with comparable RMS roughness values of 0.67 and 0.66~nm. After annealing, pronounced morphological reconstruction occurred in both films, accompanied by an increase in surface roughness to 1.29~nm for the OD-based film and 1.18~nm for the EH-based film. More importantly, distinct differences in the surface morphology developed after annealing. The OD-based film exhibited larger and more clearly defined surface domains, whereas the EH-based film showed comparatively smaller and elongated features. The results of the quantitative analysis show lateral morphological scales of approximately 180~nm for the OD-based film and 127~nm for the EH-based film. The characteristic lateral scale of the OD-based film was therefore approximately 42 \% larger than that of the EH-based film. Details of the analysis and the corresponding curves are provided in the SI section~S4 (Figure~S9). 
These side-chain-dependent differences in surface morphology may reflect differences in molecular packing and structural reorganization after annealing, which will be further examined next using GIWAXS analysis.

\subsection{Reciprocal-space analysis of annealing-driven molecular alignment}
    The side-chain-dependent differences in surface morphology may reflect differences in molecular packing and structural reorganization after annealing. To better understand this reorganization and the crystal structure of these materials, a separate batch of OD and EH samples was prepared at spin-coating speeds of 1000 rpm and 3000 rpm on SiO$_2$ (300 nm)/Si substrates for easier evaluation. The as-cast samples were then compared to those annealed with the same procedure as above. 
    Grazing-incidence wide-angle X-ray scattering (GIWAXS),  X-ray reflectometry (XRR), and grazing-incidence in-plane X-ray diffraction (IP-XRD) were used to characterize the crystal structure of the films.

\begin{figure}[h]
      \centering
      \includegraphics[width=\linewidth]{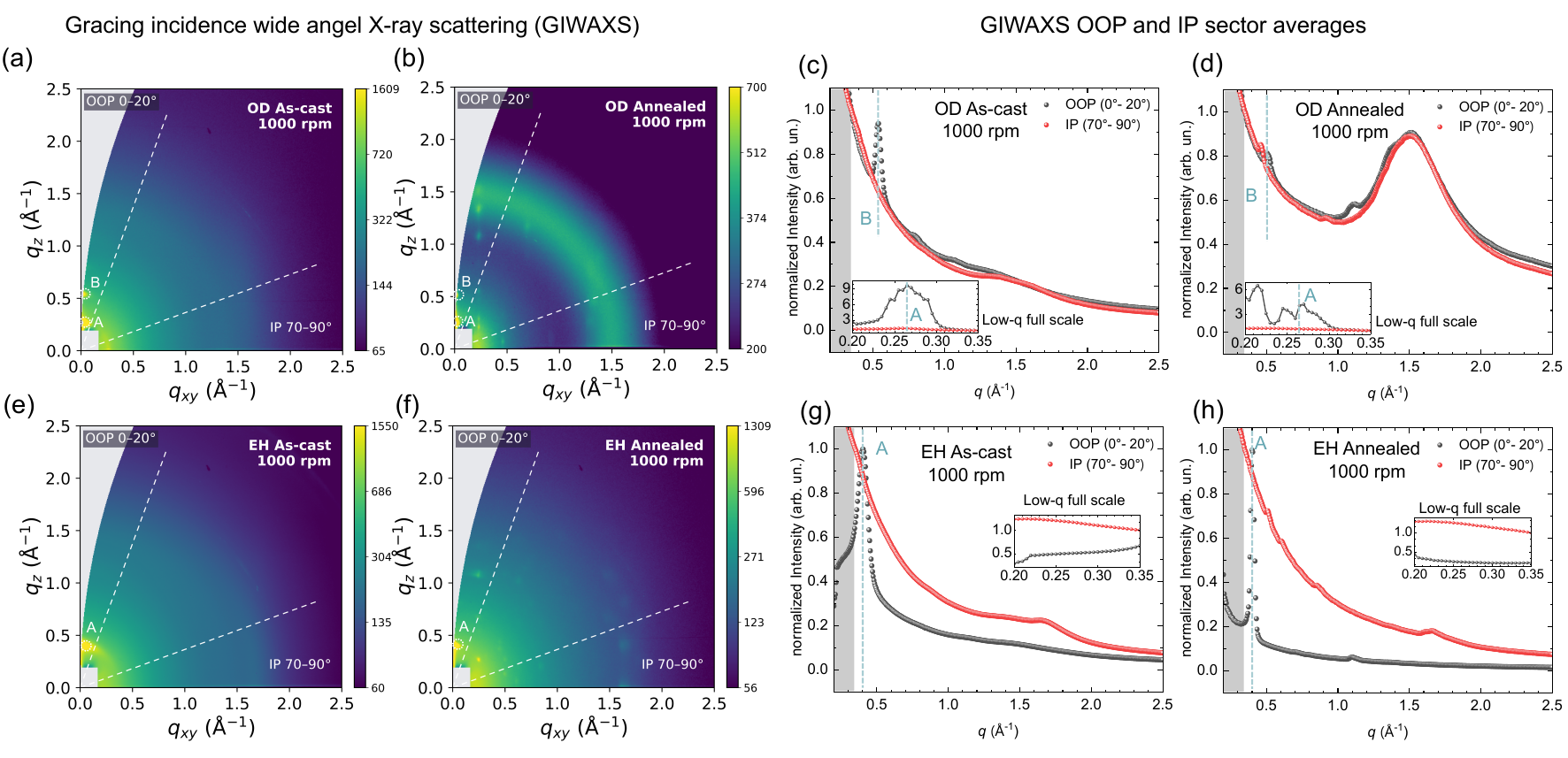}
      \caption{
        GIWAXS characterization of T2-(NDI-T2)$_2$ oligomer films.
        Two-dimensional GIWAXS patterns of OD (a,b) and EH (e,f) films prepared at 1000~rpm. Therein, (a,e) are the as-cast and (b,f) the annealed states. The grayed-out areas are inaccessible in this scan mode. The corresponding normalized, sector-averaged intensity profiles are shown in (c,d) for OD and (g,h) for EH. The out-of-plane (OOP) and in-plane (IP) sector averages cover angular ranges of 0--20$^\circ$ and 70--90$^\circ$, respectively. The insets show a zoomed-in version of the low $q$ region, indicated by the gray shading. For the GIWAXS and sector averages, the OOP lamella-like features in the first and second order are marked with A and B, respectively.
      }
      \label{GIWAX}
    \end{figure}

    Figures~\ref{GIWAX}a--d show representative GIWAXS patterns and normalized sector-averaged intensity profiles of the OD films prepared at 1000~rpm, while Figures~\ref{GIWAX}e--h present the corresponding comparison for the EH films, both before and after annealing. The sector-averaged profiles distinguish scattering in the out-of-plane (OOP, 0--20$^\circ$) and in-plane (IP, 70--90$^\circ$) sectors. Figures~S13 and S14 in the SI display all GIWAXS measurements for the OD and EH films, along with corresponding AFM images and a more detailed description.

    The as-cast OD film deposited at 1000~rpm exhibits broad, low-$q$ features, with the sector-averaged intensity profile in Figure~\ref{GIWAX}c only showing distinct peaks in the OOP direction (marked A and B), which can be attributed to the lamellar-like stacking, indicating a weakly ordered film with preferential edge-on orientation. After annealing (Figure~\ref{GIWAX}b), the OD film develops multiple additional spot-like reflections over the accessible reciprocal-space range, which causes pronounced peaks in the IP and OOP directions in the sector-averaged intensity profile (Figure~\ref{GIWAX}d). Additionally, a broad ring emerges with a radius of around 1.5~\AA$^{-1}$, which would be typical for amorphous SiO$_2$. However, this is the only sample showing this signal (see Figures~S13 and S14 in the SI), despite all having the same substrate, so the origin remains unclear.
    The 3000~rpm OD sample behaves similarly, with the spot-like features after annealing being more visible due to the lack of the amorphous ring.  
    The emergence and azimuthal localization of the reflections indicate increased crystalline order and substantial reorganization of molecular packing for both spin-coating speeds. 
    Such thermally induced changes in packing and texture are consistent with the polymorphic behavior reported for related NDI--T2 semiconductors~\cite{T2molecular}.

    For the as-cast 1000~rpm EH film, a broad OOP lamellar-like reflection is observed in the OOP direction (marked A, Figures~\ref{GIWAX}e,g). The broad reflection, in combination with the absence of any higher-order peaks, indicates a reduced crystalline coherence compared to the as-cast OD film. Its position at higher $q$ than the corresponding OD reflection indicates a smaller repeat distance. After annealing, some spot-like features appear (Figure~\ref{GIWAX}f), and the OOP reflection becomes substantially narrower (Figure~\ref{GIWAX}h). Both indicate an increase in crystalline order, as well as an increase in crystalline coherence length in the OOP direction. The IP sector-average shows only small peaks above the background after annealing, similar to the OD film when excluding the broad amorphous ring there. 
    For a spin-coating speed of 3000~rpm, the EH films behave similarly, with less pronounced spots after annealing due to reduced layer thickness.
   
    
    \begin{table}[hb]
        \centering
        \caption{Triclinic unit cell parameters extracted from the GIWAXS measurements for the annealed samples. The axial lengths $a$, $b$, and $c$, as well as the angles $\alpha$, $\beta$ and $\gamma$ are sketched in Figure~\ref{XRR}a with respect to the substrate.}
        \label{tab:GIWAXS}
        \begin{tabular}{llrrrrrrc}
            \hline
            Material & Condition & $a$ (\AA) & $b$ (\AA) & $c$ (\AA) & $\alpha$ ($^\circ$) & $\beta$ ($^\circ$) & $\gamma$ ($^\circ$) & Unit Cell Volume (\AA$^3$) \\
            \hline
            T2-(NDI-T2)$_2$-2OD & 1000 rpm & 4.89 & 27.68 & 25.29 & 78.87 & 77.23 & 82.58 & 3262 \\
            T2-(NDI-T2)$_2$-2OD & 3000 rpm & 4.90 & 27.89 & 24.33 & 79.69 & 90.24 & 77.78 & 3194 \\
            T2-(NDI-T2)$_2$-2EH & 1000 rpm & 20.91 & 26.39 & 19.28 & 57.4 & 65.86 & 73.75 & 8155 \\
            T2-(NDI-T2)$_2$-2EH & 3000 rpm & 20.64 & 26.05 & 18.95 & 58.23 & 67.21 & 74.48 & 7960 \\                            
            \hline
        \end{tabular}
    \end{table}
    
\begin{figure}[h]
      \centering
      \includegraphics[width=\linewidth]{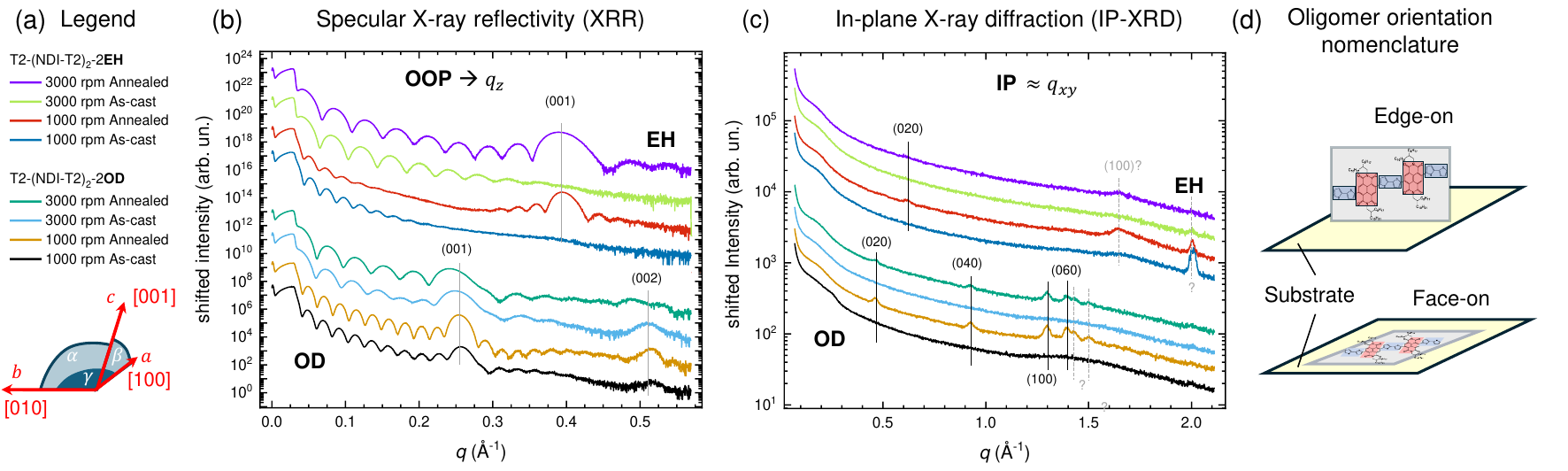}
      \caption{
         Lab-based X-ray characterization of the OD and EH films prepared at 1000 and 3000 rpm before annealing and after annealing. (a) Shared legend of (b,c), combined with nomenclature for the unit cell parameters ($a$, $b$, $c$, $\alpha$, $\beta$, and $\gamma$) and primary crystal directions ([100], [010], and [001]). 
         (b) X-ray reflectivity (XRR) curves and (c) In-plane X-ray diffraction (IP-XRD) scans. The curves are vertically offset for better comparability. Stacking directions and major peaks are assigned based on the GIWAXS unit cell measurements and literature comparison.
         (d) Illustration of the nomenclature for Edge-on and Face-on orientation of the oligomers. These do not necessarily represent the orientation of the oligomers in the unit cell. 
      }
      \label{XRR}
    \end{figure}

    For the annealed samples, the unit cell was refined from the GIWAXS measurements. Only a triclinic unit cell resulted in reasonable fits.
    The triclinic unit cell parameters, axial lengths $a$, $b$, and $c$, and angles $\alpha$, $\beta$ and $\gamma$ are collected in Table~\ref{tab:GIWAXS}, with the designation following the sketch in Figure~\ref{XRR}a. 
    
    Complementary to the GIWAXS measurements, additional lab-based XRR (Figure~\ref{XRR}b) and IP-XRD measurements (Figure~\ref{XRR}c) were performed. XRR probes in the shaded area of the GIWAXS scans along the $q_{\text{z}}$ direction at $q_{xy} = 0 $, while the IP-XRD scan provides more detailed information along the $q_{xy}$ direction for $q _{z} \approx 0$. The XRR and IP-XRD scans were performed approximately 5 and 9 months after the GIWAXS characterization, respectively, and may therefore reflect structural changes during the intervening storage periods. This limits the comparability with the GIWAXS on the fresh films, but gives some insights about the stability of these films under ambient conditions.

    All samples produce well-defined XRR signals, with Kiessig fringes corresponding to the layer thickness. The 3000~rpm films have a wider oscillation period than the 1000~rpm films, therefore showing thinner layers, which is in agreement with the results from Section~S3 in the SI. The OD samples show pronounced superlattice peaks, corresponding to the lamellar-like stacking in \ref{GIWAX}(a,b). For the EH samples, only the annealed samples show such peaks.
    Comparing the OOP periodicity extracted from GIWAXS and XRR reveals a pronounced side-chain-dependent lamellar periodicity. Both techniques show comparable spacings. The OD derivative displays a lamellar spacing of approximately 24--25~\AA, whereas the corresponding spacing of the EH derivative is approximately 15--16~\AA. This contraction is consistent with the smaller steric volume and shorter effective length of the 2-ethylhexyl substituent relative to the 2-octyldodecyl substituent, in agreement with established side-chain effects in NDI-based molecular semiconductors~\cite{jiao2019}.
    The OD films appear to retain their lamellar-like order during prolonged ambient storage, as first- and second-order lamellar-like reflections remain detectable in the subsequent XRR measurements. In contrast, among the EH films, a distinct lamellar-like reflection is observed only in the annealed samples, whereas no clearly resolved corresponding peak is detected in the as-cast samples. This difference suggests that the lamellar-like ordering in the OD films may be better preserved during storage. However, as the initial GIWAXS and subsequent XRR measurements were performed using different techniques, without a direct before-and-after comparison, the possible influence of aging cannot be conclusively established.

    As another lab-based technique, IP-XRD scans were performed to investigate the crystalline order in the film plane (along $q_{\text{xy}}$), which are shown in Figure~\ref{XRR}c. 
    Small peaks emerge over the background only for the annealed OD and EH samples, with the OD samples showing more features, indicating a better-defined crystal structure, which agrees with the GIWAXS measurements. In these samples, the 1000 rpm samples have slightly larger peaks, likely due to the increased layer thickness. A broad diffuse contribution is observed at $q \approx 1.4$~\AA$^{-1}$ in the OD films and is less pronounced for the thinner films prepared at 3000~rpm. The EH films exhibit a broad contribution at higher $q$, around 1.7~\AA$^{-1}$. The decreasing background limits an unambiguous assignment of these diffuse contributions.

   Indexing these peaks is not straightforward. From the unit cell parameters determined from GIWAXS in Table~\ref{tab:GIWAXS}, likely $d_{hkl}$ values are calculated using Equation S1 in the SI for a triclinic system. Attribution is supplemented by comparison with literature for polymeric \cite{T2struc,porzio_packing_2014,vu_third_2023,rivnay_drastic_2011} and oligomeric   \cite{T2molecular} versions of the OD system. The XRR/IP-XRD $d_{hkl}$ values are therein calculated from the peak positions using the Bragg equation in q-space (see Equation~S2 in the SI). Further details are provided in Section~S8 of the SI.

   To check the consistency between GIWAXS and XRR/IP-XRD, the first-order lattice plane distances $d_{100}$, $d_{010}$, and $d_{001}$ are calculated and compared between the GIWAXS unit cell and XRR/IP-XRD indexing in Table~\ref{GWIAXS_vs_XRR}. When higher orders were available (i.e., (020), (040), and (060)), the average lattice distance was used for the first-order value.

    \begin{table}[!h]
        \centering
        \caption{Comparison of lattice spacings $d_{h k l}$ in the first order for the annealed samples. GIWAXS values were calculated using the unit cell parameters in Table \ref{tab:GIWAXS}. XRR/IP-XRD values are based on the $(hkl)$ assignment in Figure \ref{XRR}b,c and calculated for a triclinic unit cell following equation S1 in the SI.}
        \label{GWIAXS_vs_XRR}
        \begin{tabular}{ll|cc|cc|cc}
        \hline
        \multirow{2}{*}{Material} &
        \multirow{2}{*}{Condition} &
        \multicolumn{2}{c|}{$d_{1 0 0}$ (\AA)} &
        \multicolumn{2}{c|}{$d_{0 1 0} $ (\AA)} &
        \multicolumn{2}{c}{$d_{0 0 1}$ (\AA)} \\
                            &          & GIWAXS & IP-XRD & GIWAXS & IP-XRD & GIWAXS & XRR \\ \hline
            T2-(NDI-T2)$_2$-2OD & 1000 rpm & 4.8  & 4.8    & 27.1 & 27.1 & 24.3 & 24.6 \\
            T2-(NDI-T2)$_2$-2OD & 3000 rpm & 4.8  & 4.8    & 26.8 & 27.1 & 23.9 & 24.7 \\
            T2-(NDI-T2)$_2$-2EH & 1000 rpm & 19.0 & (3.8)  & 22.2 & 20.3 & 15.4 & 15.8 \\
            T2-(NDI-T2)$_2$-2EH & 3000 rpm & 19.0 & (3.8)  & 22.1 & -    & 15.4 & 15.7 \\ \hline
        \end{tabular}
        
    \end{table}

    For the OD samples, the $d_{001}$, $d_{010}$ and $d_{100}$ agree well. Additional peaks appear between $q=1.4-1.5$~\AA{} in the IP-XRD scans, which we could not assign. This could be due to other $d_{hkl}$ or indicate other IP crystal phases.

    For the EH samples, the $d_{001}$ values obtained from GIWAXS and XRR agree closely. For the EH film prepared at 1000~rpm, the lattice spacing $d_{010}$ is 22.2~\AA{} from GIWAXS and 20.3~\AA{} from IP-XRD, corresponding to a discrepancy of approximately 8.6 \% relative to the GIWAXS value. The discrepancy for $d_{100}$ is even larger, giving 5 times smaller values for the IP-XRD scans compared to the GIWAXS measurements. The deviation in  $d_{010}$ and $d_{100}$ values could be attributed to a similar aging effect as the as-deposited EH films, although the annealed EH films seemingly only lose their IP order as a consequence of ambient storage, since the OOP $d_{001}$ distance remains consistent.

    As stated above, assigning an orientation to the molecules in the unit cell is not possible here, since their order likely differs compared to their polymer counterparts due to the much shorter backbone length. We can still use literature values, at least for the OD samples, to gauge the plausibility of the determined unit cells and $d_{hkl}$ values. To that mean, we calculate the first order lattice spacings $d_{100}$, $d_{010}$, and $d_{001}$ values for P(NDI2OD-T2) \cite{T2struc,porzio_packing_2014,vu_third_2023,rivnay_drastic_2011} and T2-(NDI-T2)$_n$-2OD oligomers \cite{T2molecular} in Table~S6 in the SI and compared them with our values from Table~\ref{GWIAXS_vs_XRR}. 
    The lamellar-like $d_{001}$ varies slightly from study to study, with the values of our oligomers (Table~\ref{GWIAXS_vs_XRR}) falling well into this range. The $d_{010} $ spacing, which is along the backbone of the polymers, should correspond to one NDI-T2 unit in the chain, which is around 14 Å for these studies \cite{T2struc,porzio_packing_2014,vu_third_2023,rivnay_drastic_2011}. The T2-(NDI-T2)$_2$-2OD oligomers here consist of more than two NDI-T2 repeat units, but $d_{010} $ is less than twice as large, indicating some tilting of the oligomers within the unit cell (i.e. a herringbone structure \cite{jung_nonfullerene_2018}).
    The $d_{100}$ distances also differ between the literature and our OD oligomers, again indicating a different IP packing compared to the polymer. 
    From this comparison, it is likely that the T2-(NDI-T2)$_2$-2OD oligomers order in an edge-on configuration. The lamellar-like stacking is comparable to the polymer, but the in-plane order differs markedly. For these OD films, annealing produces little change in the lamellar repeat distance, while the in-plane diffraction features become more clearly resolved.  
    
    Statements about the orientation of the T2-(NDI-T2)$_2$-2EH oligomers are even more limited, due to the lack of literature. Jiao et al. \cite{jiao2019} found for NDI-EH a lamellar-like stacking distance of 16.5 Å (calculated from the unit cell using Equation~S1), which is somewhat larger but still close to our $d_{001}$ distance. This indicates a more complex OOP stacking within the unit cell with an edge-on-like orientation. 

    In summary, both OD and EH show a pronounced lamellar-like stacking in the OOP direction after annealing, indicating a preferential edge-on orientation of the molecules. The lamellar period is smaller for EH due to the shorter side chain length. Both films show lamellar-like order even in the as-cast state, but IP order is established only after annealing. The precise arrangement of the molecules cannot be determined from these measurements, but differences to the corresponding polymer are to be expected based on the unit cell parameters.
    The OD oligomers on a Si/SiO$_2$ appear more stable under ambient storage, whereas the EH molecules show signs of disordering after prolonged storage. Since no direct before-and-after scans are available, statements about the long-term stability are limited here.

\subsection{Optical characterization and TDDFT simulations of the excitonic states} 

\begin{figure}[h]
  \includegraphics[width=\linewidth]{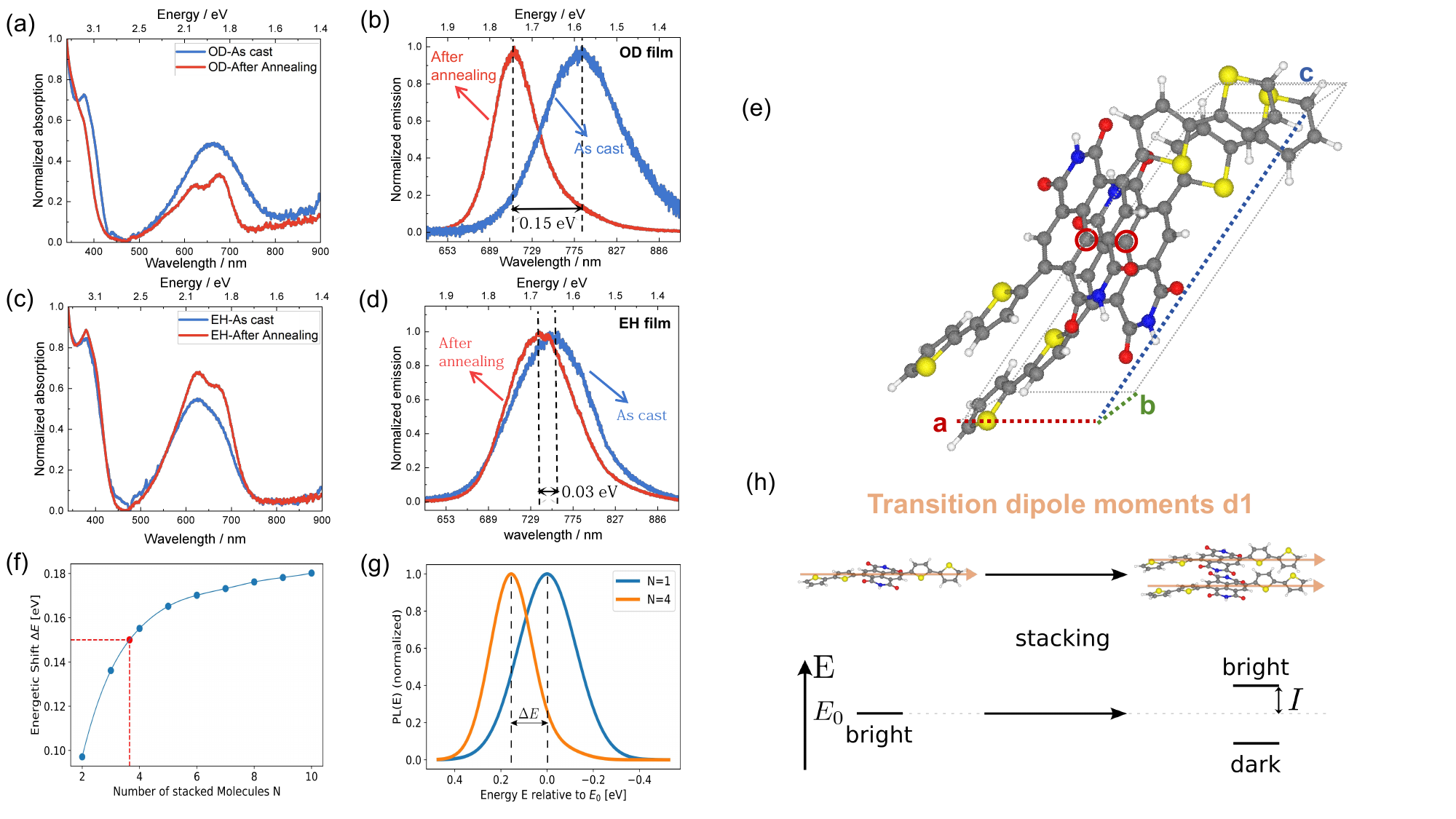}
 \caption{
(a,b) Normalized UV--vis absorption (a) and PL emission (b) spectra of the OD-based film (thickness: 10.2~nm) in the as-cast (blue) and annealed (red) states. (c,d) Corresponding normalized UV--vis absorption (c) and PL emission (d) spectra of the EH-based film (thickness: 7.6~nm). The dashed lines in (b) and (d) indicate annealing-induced PL peak shifts of approximately 0.15 and 0.03~eV, respectively. (e) Molecular arrangement used in the excitonic-coupling model, showing the relative orientation of the stacked NDI-containing molecules and the $a$, $b$, and $c$ directions. (f) Calculated excitonic energy shift, $\Delta E$, as a function of the number of stacked molecules, $N$, showing that the shift approaches saturation for $N>8$. (g) Calculated $\mathrm{PL}(E)$ spectra for $N=1$ (unaggregated reference) and $N=4$, showing an energy separation of approximately 0.15~eV. (h) Schematic illustration of transition-dipole coupling upon molecular stacking and the resulting splitting of the monomer bright state, $E_0$, into higher-energy bright and lower-energy dark exciton states.}

  \label{PL-UV}
\end{figure}

Figure~\ref{PL-UV} presents the UV–vis absorption and photoluminescence (PL) spectra of the OD- and EH-based films before and after thermal annealing, together with TDDFT calculations of the excited states in molecular dimers and a Frenkel exciton model used to describe the intermolecular excitonic interactions. Both systems exhibit blue shifts in their emission peaks after annealing (Figures~\ref{PL-UV}b,d), but the magnitude shows a significant dependence on the side chain. The long side chain OD system exhibits a blue shift of approximately 0.15~eV, while the short side chain EH system shows a shift of only about 0.03~eV. Such annealing-induced PL blue shifts can be associated with enhanced intermolecular excitonic coupling in more closely packed $\pi$-conjugated aggregates. In idealized H-aggregate models, cofacial stacking leads to positive intermolecular excitonic coupling, which can shift the optically allowed transition toward higher energy~\cite{PLshift01}. 

This pronounced difference in the PL shifts indicates that thermal annealing affects the local excitonic aggregate configurations of the two materials differently. For the OD-based films, the substantial blue shift suggests a pronounced reorganization from relatively weakly coupled or disordered aggregates toward more strongly coupled H-type $\pi$-stacked aggregates. By contrast, the much smaller PL shift of the EH-based films suggests that annealing causes a comparatively limited change in the local intermolecular environment governing the emission. 

The UV--Vis absorption spectra also reveal side-chain-dependent responses to annealing (Figures~\ref{PL-UV}a,c). For the OD-based film, annealing decreases the relative intensity and alters the line shape of the broad low-energy absorption band between approximately 600 and 750~nm (Figure~\ref{PL-UV}a). Together with the pronounced PL blue shift discussed above, this change is consistent with a substantial annealing-induced reorganization of the local excitonic environment and a modification of intermolecular excitonic coupling~\cite{PLshift02,absorption01}. 
By contrast, the corresponding low-energy absorption band becomes somewhat more pronounced relative to the high-energy band in the annealed EH-based film (Figure~\ref{PL-UV}c). 

To investigate whether intermolecular excitonic coupling can account for the observed blue shift, TDDFT calculations on a $\pi$-stacked molecular dimer were combined with a finite-stack Frenkel-exciton model (Figures~\ref{PL-UV}e--h; full details in the Methods section). The alkyl side chains were replaced by hydrogen atoms to focus on interactions between the conjugated backbones. After geometry optimization at the DFT-D3BJ level, the closest intermolecular nuclear separation across the van der Waals gap was 3.716~\AA{} (red circles in Figure~\ref{PL-UV}e), consistent with close $\pi$--$\pi$ stacking. The TDDFT-derived nearest-neighbour excitonic coupling is positive, $I = 0.0966$~eV. Within the dimer model, this coupling produces an upper bright state at $E_0 + I$ and a lower dark state at $E_0 - I$, corresponding to a total bright--dark splitting of $2I = 0.1932$~eV and H-type excitonic coupling (Figure~\ref{PL-UV}h). Extending the model to finite stacks produces collective Frenkel-exciton states distributed over multiple molecules~\cite{X4}. Phenomenological PL model spectra were constructed by Gaussian broadening of the excitonic transitions weighted by their squared transition dipole moments. The peak shift, $\Delta E$, is defined relative to the $N = 1$ unaggregated reference spectrum. This shift increases with $N$, reaching approximately 0.18~eV for the largest stacks considered in Figure~\ref{PL-UV}f. For $N = 4$, the phenomenological model spectrum exhibits a peak shift of approximately 0.15~eV relative to the $N = 1$ reference (Figure~\ref{PL-UV}g), comparable to the annealing-induced PL blue shift observed for the OD-based film.

Taking $N = 1$ as an idealized reference for the weakly coupled as-cast state, the observed OD blue shift is consistent, within this model, with an effective stack of approximately three to four coupled molecules after annealing. This estimate depends on the assumed reference state and model parameters and does not uniquely determine the physical aggregate size or exciton coherence length. The much smaller PL shift of the EH-based film indicates a smaller annealing-induced change in the emissive environment, but does not exclude $\pi$-stacking in either state. Because the side chains were omitted from the calculations, the model illustrates the effect of backbone aggregation rather than directly predicting the difference between the OD and EH films.

Overall, the absorption and PL results indicate that annealing affects the electronic and excitonic environments of the two materials differently. While the AFM results show a morphological reconstruction for both materials, the OD-based film shows a greater increase in grain size and a larger characteristic lateral morphological scale. Together, the optical measurements and model analysis support a more pronounced annealing-induced change in the local excitonic coupling environment of the OD-based films than of the EH-based films.

\subsection{Side-chain-dependent vertical charge transport}

\begin{figure}[hbt]
  \centering
  \includegraphics[width=\linewidth]{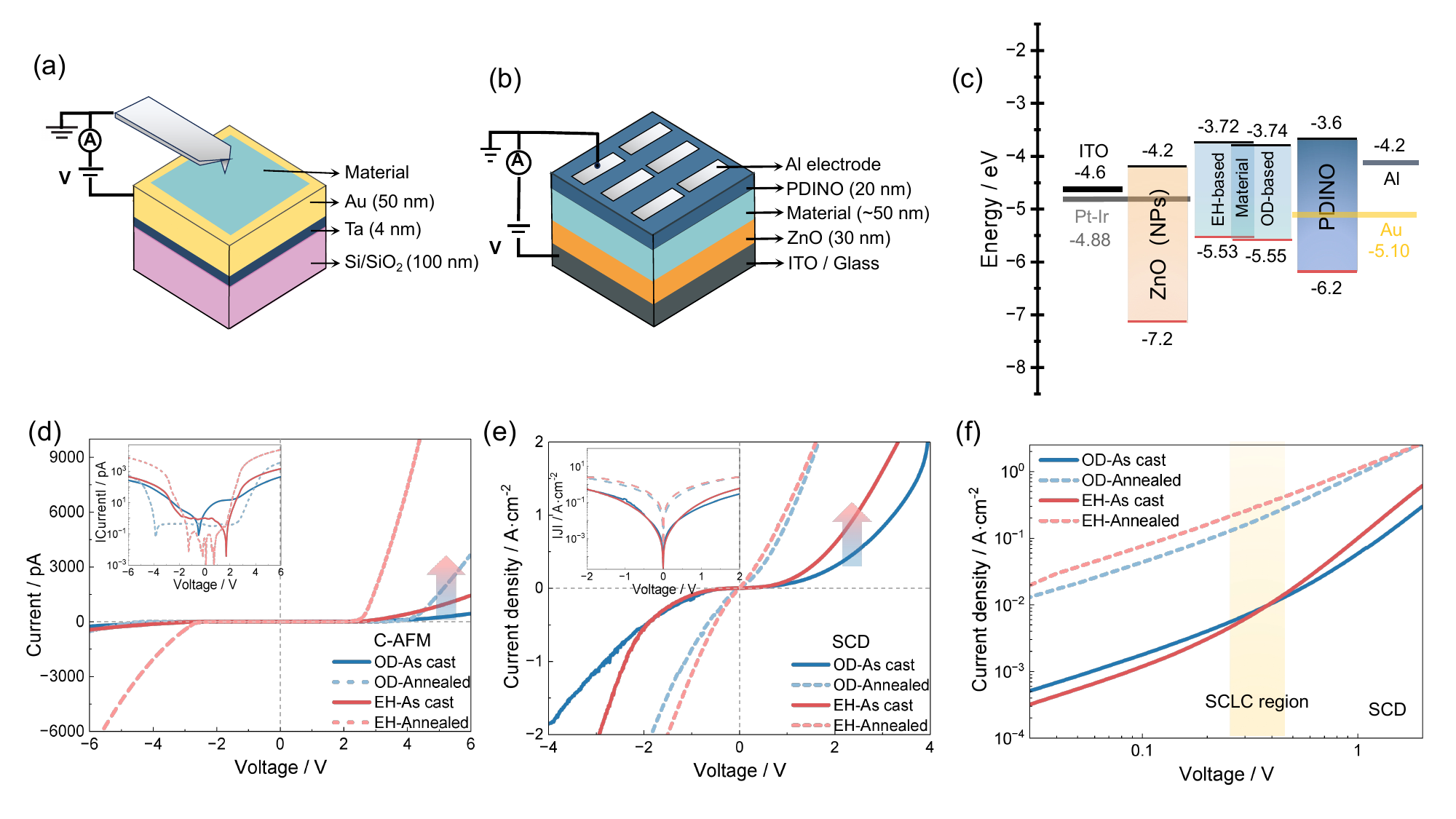}
  \caption{Vertical charge-transport characterization of the NDI-based films.
(a) Schematic illustration of the conductive atomic force microscopy (C-AFM) measurement configuration using a Pt-coated conductive tip in direct contact with the NDI-based film.
(b) Schematic of the electron-only device used for single-carrier-device (SCD) measurements, where SCD stands for single-carrier device.
(c) Energy-level diagram of the two NDI-based materials together with the electrode and interlayer materials used in the C-AFM and electron-only device measurements. The Pt--Ir tip and Au substrate used for C-AFM are highlighted in gray and yellow, respectively, whereas ITO/ZnO and PDINO/Al define the contacts in the electron-only devices.
(d) C-AFM current--voltage ($I$--$V$) characteristics of OD and EH films in the as-cast and annealed states; the inset shows an enlarged view of the low-bias region.
(e) Current density--voltage ($J$--$V$) characteristics of the corresponding SCDs; the inset shows an enlarged view of the low-voltage region.
(f) Double-logarithmic $J$--$V$ plots highlighting the low-field linear and high-field space-charge-limited-current (SCLC) regimes. Solid lines represent annealed films, whereas dashed lines represent as-cast films.}
  \label{verticaltransport}
\end{figure}

The vertical charge-transport properties of the NDI-based films were first investigated locally using conductive atomic force microscopy (C-AFM). A Pt--Ir-coated conductive tip with a nominal apex radius of approximately 35~nm was used as the top nanoelectrode and brought into direct contact with the organic film deposited on an Au/Ta/SiO$_2$/Si substrate (Figure~\ref{verticaltransport}a). Compared with the planar electrode structures, this point-contact geometry confines the current path predominantly perpendicular to the substrate plane and reduces contributions from lateral leakage and edge effects~\cite{leak}.
The corresponding energy-level alignment is shown in Figure~\ref{verticaltransport}c. The work functions of the Pt--Ir tip and Au electrode were taken from previously reported values~\cite{PtIrWF,AuWF}, whereas the HOMO and LUMO energies of the NDI-based material were obtained from Ref.~\cite{T2molecular}. $J$--$V$ characteristics of the films deposited at 3000~rpm are presented in Figure~\ref{verticaltransport}d. The complete C-AFM datasets are provided in Section S5 and Figures S10 and S11. For both materials, annealing increased the current density. The EH-based film exhibited a higher current density than the OD-based film, especially after annealing. The inset of Figure~\ref{verticaltransport}d replots the absolute current density, $|J|$, on a semilogarithmic scale. Notably, the current magnitudes differ between positive and negative biases of equal magnitude. The asymmetric response is consistent with non-equivalent contact conditions at the Pt–Ir/oligomer and oligomer/Au interfaces. The nominal energy-level offsets provide a qualitative basis for this interpretation. Because the measured current can additionally be influenced by the local tip--sample contact conditions and topographical inhomogeneity of the films, the C-AFM results are primarily used for a qualitative comparative assessment of local vertical charge transport, rather than for direct quantitative analysis~\cite{gicevivcius2025,reid2008,fernando2023}.

To further investigate the vertical electron transport over the film, electron-only devices with the structure ITO/ZnO/NDI-based material/PDINO/Al were fabricated (Figure~\ref{verticaltransport}b). The organic active layers were prepared independently from the ultrathin films used for the C-AFM measurements, and the Al top electrodes were defined through a shadow mask, giving an effective active area of 0.04~cm$^2$ for each device pixel. ZnO and PDINO were used as electron-selective interlayers because they provide more favorable energy-level alignment for electron injection than direct contact with high-work-function metals such as Au or Pt--Ir, while suppressing hole injection~\cite{hau2008,zhang2014}. The representative linear $J$--$V$ curves of the 3000~rpm devices are shown in Figure~\ref{verticaltransport}e. Both materials exhibited nearly symmetric responses under forward and reverse bias, indicating relatively balanced electron injection from the two electrodes (Figure~\ref{verticaltransport}e inset). Annealing clearly increased the current density of both devices, consistent with the C-AFM results. The complete electron-only-device characteristics are provided in Section S6 and Figure S12, Supporting Information, and the extracted conductivities and mobilities are listed in Tables S4 and S5, respectively.

The $J$--$V$ characteristics were further examined on a double-logarithmic scale (Figure~\ref{verticaltransport}f). The local derivative $m=\mathrm{d}(\log J)/\mathrm{d}(\log V)$, where $J$ is the current density and $V$ is the applied voltage, was calculated to identify the voltage range in which $m$ lies near~2, indicating entry into the space-charge-limited current (SCLC) regime described by the Mott--Gurney law~\cite{MGlaw,SCLC0001,SCLC0002}. At higher voltages, values of $m>2$ may arise from trap filling, field-dependent mobility, or residual contact-related effects~\cite{MGlaw,SCLC0001,SCLC0003,SCLC0004}. The electron mobility was extracted from the identified SCLC region using $J=(9/8)\varepsilon_0\varepsilon_{\mathrm{r}}\mu V^2/L^3$, where $\varepsilon_0$ is the vacuum permittivity, $\varepsilon_{\mathrm{r}}$ is the relative permittivity of the organic semiconductor, $\mu$ is the electron mobility, and $L$ is the organic-layer thickness. For the 3000~rpm devices, organic-layer thicknesses of 30 and 34~nm and relative permittivities of 3.0 and 3.2 were used for the OD- and EH-based films, respectively. These values are consistent with the experimentally reported relative permittivity of approximately 3.0 for pristine P(NDI2OD-T2)~\cite{politano2023,fujiwara2007,yang2025}. The effective vertical conductivity was extracted at $V=0$ using $\sigma=L\left.\mathrm{d}J/\mathrm{d}V\right|_{V=0}$. The resulting mobility and conductivity values are summarized in Table~\ref{transportparameters}.

\begin{table}[h]
\centering
\caption{Electron mobility ($\mu$) and conductivity ($\sigma$) of the NDI-based films prepared at 3000~rpm.}
\begin{tabular}{lcccc}
\hline
Sample
& \multicolumn{2}{c}{Electron mobility, $\mu$}
& \multicolumn{2}{c}{Conductivity, $\sigma$} \\
& As-cast & Annealed & As-cast & Annealed \\
& \multicolumn{2}{c}{$(\mathrm{cm^2\,V^{-1}\,s^{-1}})$}
& \multicolumn{2}{c}{$(\mathrm{S\,cm^{-1}})$} \\
\hline
OD-based film
& $5.3\times10^{-6}$ & $4.9\times10^{-5}$
& $4.15\times10^{-8}$ & $2.56\times10^{-7}$ \\
EH-based film
& $1.2\times10^{-5}$ & $5.1\times10^{-5}$
& $2.18\times10^{-8}$ & $2.94\times10^{-7}$ \\
\hline
\end{tabular}
\label{transportparameters}
\end{table}

Overall, annealing improved both the electron mobility and conductivity of both materials. For the films prepared at 3000 rpm, the mobility increased by approximately 9.2-fold for the OD-based film and 4.3-fold for the EH-based film. The two materials therefore reached comparable mobilities after annealing, with the EH-based film exhibiting a slightly higher value. The corresponding conductivity increased by approximately 6.2-fold and 13.5-fold for the OD- and EH-based films, respectively, indicating a larger conductivity enhancement for the EH-based film.
The conductivity can be expressed as $\sigma=en\mu$, where $e$ is the elementary charge, $n$ is the concentration of injected carriers, and $\mu$ is the electron mobility. Since the OD- and EH-based films have comparable thicknesses and relative permittivities, their injected carrier concentrations are expected to be similar under comparable injection conditions~\cite{Wetzelaer2014}. Accordingly, the higher conductivity of the annealed EH-based film is consistent with its slightly higher electron mobility. Therefore, the higher conductivity of the EH-based film is consistent with its higher extracted electron mobility, further indicating more efficient vertical electron transport.

The structural characterization presented above suggests a preferential edge-on alignment of both materials after annealing. Owing to its shorter side chains, the EH-based film exhibits a smaller lamellar stacking distance than the OD-based film. This shorter distance may facilitate charge transport along the out-of-plane lamellar-stacking direction and thereby account for the higher mobility and conductivity of the EH-based film. By contrast, the IP-XRD measurements show more pronounced crystalline features for the OD-based film. The relative contribution of in-plane order therefore appears to be smaller for the present out-of-plane transport geometry.

Taken together, the optical measurements and exciton calculations relate the annealing-induced spectral changes to modifications in intermolecular excitonic coupling, while the electrical measurements demonstrate enhanced vertical charge transport. Both responses are sensitive to molecular packing~\cite{vu2023,Noriega2013,Troisi2006,Steyrleuthner2014}. The larger PL blue shift and mobility enhancement of the OD-based film are qualitatively consistent with its more pronounced annealing-induced molecular reorganization, accompanied by changes in $\pi$--$\pi$ packing and pronounced lamellar ordering. Nevertheless, the EH-based films exhibit consistently higher absolute electron mobilities and conductivities after annealing despite their less well-defined in-plane order. These results suggest that out-of-plane lamellar packing and stacking distance are more important for vertical charge transport than the degree of in-plane order.

    \section{Conclusion}
    
    In this work, we investigated the influence of branched side-chain structure on film morphology, molecular packing, optical response, and vertical electron transport in T2-(NDI-T2)$_2$-based n-type conjugated oligomers. By comparing OD- and EH-based derivatives with the same conjugated backbone, we established a direct correlation between side-chain architecture, annealing-induced structural evolution, and out-of-plane charge transport.
    
    AFM and thermal analysis show that the two side-chain systems respond differently to thermal treatment. Combined GIWAXS and XRR/IP-XRD analysis revealed pronounced lamellar stacking, suggesting an edge-on orientation. After annealing, the lamellar order becomes more pronounced and in-plane order appears, which was mostly absent in the as-cast state. 
    Optical spectroscopy further reveals distinct side-chain-dependent aggregation behavior. The OD-based film exhibits a larger annealing-induced PL blue shift, indicating a greater change in effective excitonic aggregate configuration after thermal treatment. In contrast, the smaller PL shift of the EH-based film suggests that its interchain coupling environment is already established in the as-cast state. TDDFT calculations and Frenkel exciton analysis support this interpretation by showing that molecular $\pi$--$\pi$ stacking can induce excitonic splitting and a blue shift of the optical transition. C-AFM and electron-only vertical diode measurements further demonstrate that annealing improves vertical electron transport in both systems. The $J$--$V$ characteristics show increased current density after annealing, and the double-logarithmic analysis indicates an entry into the space-charge-limited current (SCLC) regime within the selected voltage range. Quantitative analysis based on the Mott--Gurney relation shows that annealing improves the electron mobility and conductivity of both materials. The EH-based films exhibit higher values at all spin-coating speeds, reaching a maximum mobility of $1.2\times10^{-4}$~cm$^2$~V$^{-1}$~s$^{-1}$. The conductivity extracted from the linear region also increases after annealing. Since $\sigma$ reflects the combined contribution of injected carrier concentration and mobility, the consistently higher conductivity of the EH-based films is consistent with their higher extracted electron mobility.
    Overall, these results show that side-chain structure controls the balance between molecular rearrangement, packing order, excitonic coupling, and vertical electron mobility in NDI-based oligomers.
    
    This study suggests that side-chain engineering is an effective strategy for tuning out-of-plane charge transport in n-type organic semiconductors. In the present T2-(NDI-T2)$_2$ oligomer system, the EH side chain provides more favorable vertical electron transport than the OD side chain after thermal annealing. These findings provide guidance for the molecular design of n-type organic semiconductor materials for vertical optoelectronic devices.

\section{Materials and Methods}

\subsection{Material synthesis}

T2-(NDI-T2)$_2$-2OD was prepared following the procedure reported by Matsidik et al., who also provided its structural characterization, including $^{1}$H and $^{13}$C NMR data~\cite{T2molecular}. The corresponding oligomer bearing shorter 2-ethylhexyl side chains, T2-(NDI-T2)$_2$-2EH, was synthesized using an adapted palladium-catalyzed direct-arylation procedure.
Briefly, bithiophene (5~equiv) and NDI--Br$_2$ (1~equiv) were reacted in degassed mesitylene containing K$_2$CO$_3$, pivalic acid, and a catalytic amount of Pd$_2$(dba)$_3$ under an argon atmosphere. The mixture was heated at 90~$^\circ$C for 24~h, cooled to room temperature, and purified by column chromatography to yield the target compound. The structure of the EH derivative was characterized by $^{1}$H and $^{13}$C NMR spectroscopy. Synthetic details and NMR data for the EH derivative are provided in the SI.

\subsection{Thin film preparation}
Thin films were prepared on two types of substrates: 300\, nm thermally grown SiO$_2$/Si substrates (used for GIWAXS and XRR measurements) and Au-coated Si substrates (used for AFM and conductivity-related characterizations). Both substrates were cut to dimensions of 0.7\,$\times$\,0.7\,cm$^2$. The Au-coated substrates with the layer stack
$\mathrm{Au}\,(50~\mathrm{nm})/\mathrm{Ta}\,(4~\mathrm{nm})/
\mathrm{SiO_2}\,(100~\mathrm{nm})/\mathrm{Si}$
were fabricated by radio-frequency (RF) sputtering. The samples used for XRR and GIWAXS measurements were prepared in the same batch under identical processing conditions.

Before deposition, all substrates were cleaned by successive sonication in acetone, 2-propanol, and deionized water and dried under a nitrogen gas flow. Oligomer solutions were dynamically spin-coated onto the cleaned substrates at room temperature. Spin speeds range from 1000 to 5000\, rpm, producing uniform thin films. 

Thermal annealing was performed using a Linkam THMS600 thermal stage in a nitrogen atmosphere to promote molecular rearrangement while reducing oxidative degradation. To achieve controlled crystallization, the annealing temperatures were carefully selected to be slightly above the crystallization temperature ($T_\mathrm{c}$) of each material. 145~$^\circ$C for T2-(NDI-T2)$_2$-2OD ($T_\mathrm{c}$ = 133.8~$^\circ$C) and 208~$^\circ$C for T2-(NDI-T2)$_2$-2EH ($T_\mathrm{c}$ = 204.92~$^\circ$C).

To reduce interfacial stress arising from the thermal-expansion mismatch between the substrate and the organic layer, the heating rate was controlled at 10~K~min$^{-1}$ above 80~$^\circ$C and reduced to below 2~K~min$^{-1}$ as the target annealing temperature was approached. Preliminary tests using annealing times of 10, 30, and 60~min were performed to evaluate the effect of annealing duration on the film morphology. Based on these tests, an annealing time of 30~min was selected for the subsequent structural, optical, and electrical measurements. This protocol ensured consistent thermal treatment when comparing the as-cast and annealed OD- and EH-based films.

\subsection{GIWAXS for molecular packing analysis} 

Grazing-incidence wide-angle X-ray scattering (GIWAXS) was conducted at BL11 - NCD-SWEET at the ALBA synchrotron using a monochromatic X-ray beam (12.4~keV) with an incidence angle of $0.12^{\circ}$. The beam size was $30\,\mu\mathrm{m}$ (vertical) by $130\,\mu\mathrm{m}$ (horizontal), and diffraction patterns were collected using a Rayonix LX 255 HS detector positioned 151.6~mm from the sample. The sample-to-detector distance and beam position on the detector were verified using a Cr$_2$O$_3$ standard. Each dataset was acquired with exposure times ranging from 1 to 20~s. The two-dimensional GIWAXS patterns were processed using GIWAXS-tools. One-dimensional out-of-plane (OOP) and in-plane (IP) intensity profiles were obtained by azimuthal integration over angular ranges of $0^{\circ}$--$20^{\circ}$ and $70^{\circ}$--$90^{\circ}$ relative to the sample surface normal, respectively, and subsequently normalized for comparison.

\subsection{X-ray reflectivity and in-plane X-ray diffraction} 

All X-ray measurements were made with a Rigaku SmartLab diffractometer equipped with a 9 kW rotating anode and Cu K$_{\alpha}$ radiation in a parallel-beam configuration. 
A K$_{\beta}$ filter was inserted into the incident beam path during the XRR measurements to suppress Cu K$_{\beta}$ radiation and improve spectral purity. 
In-plane X-ray diffraction (IP-XRD) scans were performed using the in-plane arm option of the Rigaku tool without a spectral filter to maximize intensity. The detector was moved in $2\theta_\chi$ in the film plane, while the sample rotated by $\varphi$, with a constant incidence angle of 0.19° and detector angle of 0° for all measurements. 

\subsection{Steady-state optical measurements} 

PL spectra were measured using a Horiba XploRA Plus Raman microscope equipped with a 100$\times$, 0.9~NA objective. A 532~nm diode-pumped solid-state (DPSS) continuous wave (CW) laser served as the excitation source, with an excitation power of 100~$\mu$W measured under the objective. The spectrometer was configured with a 600~L/mm grating and an electron-multiplying charge-coupled device (EMCCD) detector. The lateral optical resolution under this configuration is $\sim$0.7~$\mu$m, which is sufficient to resolve individual terraces on the sample surface (typical lateral dimensions of 1-5~$\mu$m).

Thin-film absorption spectra were collected using a Flame-S UV-Vis spectrometer (Ocean Optics) with an integration time of 3~ms and averaging over 666~scans, controlled via OceanView~1.5.2 software.

\subsection*{Spectroscopic ellipsometry for optical characterization}

Spectroscopic ellipsometry was performed using a J.A. Woollam M-2000 system to characterize independently prepared T2-(NDI-T2)$_2$-2OD and T2-(NDI-T2)$_2$-2EH thin films deposited on Si/SiO$_2$ and Au substrates. These ellipsometry samples were distinct from the sample batch used for the XRR and GIWAXS measurements. Data were collected at seven incident angles (45$^\circ$--75$^\circ$) across the 245--1700~nm range, with a lateral resolution of 200--300~$\mu$m using focusing probes.

Film thickness was first extracted from the transparent spectral region ($<$1.5~eV) using a Cauchy model:

\[
n(\lambda) = A + \frac{B}{\lambda^2} + \frac{C}{\lambda^4}
\]

achieving a fitting precision better than 0.1~nm. This thickness value was then fixed during subsequent modeling of the absorbing region.

The complex dielectric function ($\varepsilon = \varepsilon_1 + i\varepsilon_2$) was parameterized using a hybrid oscillator model combining Cody–Lorentz oscillators (to capture $\pi$–$\pi^*$ and charge-transfer transitions) and Gaussian oscillators (to model annealing-induced sub-gap or vibronic features). Surface roughness values extracted from AFM measurements were used as initial inputs, and an effective medium approximation (EMA) layer composed of 50\% film material and 50\% voids was applied to account for morphological effects. All analysis was conducted using \textit{CompleteEASE} software. The effective relative permittivities used in the SCLC analysis were obtained from the real part of the ellipsometry-derived dielectric functions. Values of $\varepsilon_r=3.0$ and 3.2 were used for the OD- and EH-based organic layers, respectively.

\subsection{Computational methods and exciton modeling}

\subsubsection{Electronic-structure calculations}

All electronic-structure calculations were performed using CP2K~\cite{Kuhne2020CP2K}. To reduce the computational cost, the alkyl side chains were truncated and replaced by hydrogens, keeping the conjugated molecular backbone intact. Geometry optimizations in the singlet ground state were performed using the PBE exchange-correlation functional~\cite{Perdew1996PBE} together with the DFT-D3 dispersion correction and Becke--Johnson damping~\cite{Grimme2011D3BJ}. The TZV2P-GTH basis set and GTH-PBE pseudopotentials were employed~\cite{Goedecker1996GTH,Hartwigsen1998GTH,Krack2005GTH}. The initial dimer geometry consisted of two approximately cofacial molecules with slightly displaced NDI cores. This optimized geometry was subsequently used for the excited-state calculations.

\subsubsection{Excited-state calculations}

Vertical singlet excitation energies and transition dipole moments were calculated using linear-response time-dependent density functional theory (TDDFT)~\cite{Runge1984TDDFT,Casida1998HighLying,Casida1998ExcitedState,Vasiliev1999TDLDA}. The calculations employed the LC-PBEOP functional~\cite{Tawada2004LC}, a long-range-corrected version of the PBE functional, with a range-separation parameter of $0.33~\mathrm{Bohr}^{-1}$. The lowest singlet excited states were calculated using the Tamm--Dancoff approximation~\cite{Benedict1998Optical}. The lowest relevant monomer excitation had a transition dipole of $d_{1}=-3.974\,e a_{0}$.

\subsubsection{Frenkel-exciton model}

The dependence of the excitonic states on aggregate size was investigated using a one-dimensional Frenkel-exciton model containing $N$ identical molecules. Each molecule contributed one local excited state with the excitation energy $E_{0}$. Only coupling between directly adjacent molecules was included. The Hamiltonian was defined as

\begin{equation}
\hat{H}_{N}
=
E_{0}\sum_{i=1}^{N}\lvert i\rangle\langle i\rvert
+
I\sum_{i=1}^{N-1}
\left(
\lvert i\rangle\langle i+1\rvert
+
\lvert i+1\rangle\langle i\rvert
\right).
\end{equation}

Here, $\lvert i\rangle$ denotes a state in which molecule $i$ is electronically excited and all other molecules remain in their ground states. For a dimer with $I>0$ and approximately parallel molecular transition dipoles, diagonalization of the Hamiltonian gives the bright and dark excitonic states with energies $E_{\mathrm{bright}}=E_{0}+I$ and $E_{\mathrm{dark}}=E_{0}-I$, corresponding to H-type excitonic coupling. The Hamiltonian was diagonalized, and $c_{ik}$ denotes the contribution of the local excitation on molecule $i$ to excitonic eigenstate $k$. The transition dipole moment of state $k$ was calculated as

\begin{equation}
D_{k}
=
d_{1}\sum_{i=1}^{N}c_{ik}.
\end{equation}

A phenomenologically broadened model spectrum was constructed from the calculated excitonic energies and transition dipole moments according to

\begin{equation}
S(E)
=
C\sum_{k=1}^{N}\lvert D_{k}\rvert^{2}
\exp\left[
-\frac{\left(E-E_{k}\right)^{2}}{2\sigma^{2}}
\right].
\end{equation}

Here, $E$ is the photon energy, $E_{k}$ is the energy of excitonic state $k$, $D_{k}$ is its transition dipole moment, $\sigma$ is the Gaussian broadening parameter, and $C$ is a normalization constant. The broadening parameter was set to $\sigma=90~\mathrm{meV}$ and $125~\mathrm{meV}$ for $N=1$ and $N=4$, respectively.

\subsection{AFM based morphology and local conductivity probing}
AFM and C-AFM measurements were conducted using a SmartSPM\texttrademark~1000 system (AIST-NT). Standard AFM imaging was performed under ambient conditions with controlled temperature and humidity using a commercially available NSG10 cantilever with the tip radius of 10\,nm.

For C-AFM, the system operates in contact mode under a nitrogen atmosphere while minimizing environmental interference such as humidity, ambient light, and acoustic noise. A Pt-Ir alloy-coated NSG10 tip (TipsNano) with a nominal radius of approximately 35\,nm was used for C-AFM measurements. The active semiconducting layer was spin-coated onto a gold-coated substrate to ensure reliable electrical contact.

During scanning, a constant sample bias of 0.1\,V was applied, and the resulting current was collected via the integrated transimpedance amplifier. Before the measurement, the electrical functionality of the setup was verified by acquiring I-V curves on freshly cleaved highly ordered pyrolytic graphite (HOPG). To protect both the tip and the sample, a current compliance limit of 100\,nA was enforced. All measurements were repeated across multiple sample areas to verify the reproducibility of the results.

\subsection{Vertical diode fabrication and I–V measurement}
Vertical device architectures with the configuration ITO/ZnO/active layer/PDINO/Al were fabricated on ITO-coated glass substrates. Substrates were sequentially cleaned by ultrasonication in acetone, deionized water, and isopropanol for 10~min each, followed by nitrogen blow-drying and UV-ozone treatment for 15~min to enhance surface wettability. 

A ZnO electron transport layer ($\sim$30~nm) was deposited via spin-coating from an isopropanol-based ZnO precursor solution at 2500~rpm for 60~s. The semiconducting layers were spin-coated at 1000, 3000, and 5000~rpm and subsequently dried. Before deposition of the upper interfacial and electrode layers, the organic-layer thickness was measured by spectroscopic ellipsometry. Subsequently, a 20~nm interfacial layer of PDINO was deposited by spin-coating a methanol solution (1~mg~mL$^{-1}$) at 4000~rpm for 60~s. 

Aluminum top electrodes ($\sim$100~nm) were thermally evaporated through a shadow mask under high vacuum (3~$\times$~10$^{-4}$~Pa), defining six pixels per substrate with an effective device area of 0.04~cm$^{2}$. All device fabrication steps were carried out in a nitrogen-filled glovebox. 

The as-cast current density--voltage ($J$--$V$) characteristics were first measured using a Keithley 236 SourceMeter under nitrogen. The completed devices were subsequently annealed in situ for 30~min at 145~$^\circ$C for the OD-based devices and 208~$^\circ$C for the EH-based devices. After cooling to room temperature, the $J$--$V$ characteristics were measured again under the same conditions. Because the organic layer was no longer accessible after completion of the device stack, the thickness measured before device completion was used in the SCLC analysis of both the as-cast and annealed devices.


\medskip
\textbf{Supporting Information} \par 

\medskip
\textbf{Acknowledgements} \par 
Funded by the Deutsche Forschungsgemeinschaft (DFG, German Research Foundation) – TRR 386 – TP B5, Project number 514664767.

\section*{Conflict of interest}
The authors declare no conflict of interest.

\medskip

%

\bibliographystyle{MSP}
\bibliography{references}




\end{document}